# Morphology-engineered nanostructured silver- and antimony-telluride films for flexible thermoelectric generators

Ankit Kashyap[a,b], Conner Wallace[b], Geetu Sharma[b], Collin Rowe[b], Mahima Sasikumar[b], Niraj Kumar Singh[d], Per Eklund[d,e], Theodorian Borca-Tasciuc[c], Ganpati Ramanath[b,d,e*], and Ajay Soni[a*]

[a]School of Physical Sciences, Indian Institute of Technology Mandi, 175005, Himachal Pradesh, India.

[b]Rensselaer Polytechnic Institute, Materials Science & Engineering Department, Troy, NY 12180, USA.

[c]Rensselaer Polytechnic Institute, Department of Mechanical Engineering, Troy, NY 12180, USA.

[d]Inorganic Chemistry, Ångström Laboratory, Uppsala University, SE-751 21 Uppsala, Sweden.

[e]Wallenberg Initiative Materials Science for Sustainability (WISE), Uppsala University, Sweden.

*Corresponding authors: ganapr@rpi.edu, ganpati.ramanath@kemi.uu.se and ajay@iitmandi.ac.in

**Abstract**

Harvesting low-grade heat to electricity is attractive for powering wearable electronic devices. Here, we demonstrate nW-scale thermoelectric power generation in devices from thin film assemblies of microwave-synthesized p-$Sb_2Te_3$ nanoplates and n-$Ag_2Te$ nanowires on polyvinylidene fluoride (PVDF) membranes. While microwave cycling is crucial for $Ag_2Te$ nanocrystal shaping, $Sb_2Te_3$ formation is sensitive to precursors and surfactant concentrations. Introducing S doping in $Sb_2Te_3$ in the $1 \leq x_S \leq 1.5$ at.% range via thioglycolic acid during synthesis yields an up to eightfold higher power-factor, due to a fivefold increase in electrical conductivity and ~25% increase in Seebeck coefficient. Our microfilm devices generate up to 33.6 mV from 5 °C $\leq \Delta T \leq$ 50 °C thermal gradients, with ~120 nW maximum power output at $\Delta T$ = 30 °C, which is sixtyfold higher than $Sb_2Te_3$ paper devices. Mechanical bending can increase device resistance by up to ~125% due to diminished inter-nanostructure electronic transport. These findings provide insights for integrating synthesis, morphology engineering and device design for next-generation wearable thermoelectric systems.

## 1. Introduction

Wearable electronic devices for diverse applications[1] in physiological monitoring[2], sensing[3], and diagnostics, are typically powered by micro-/milli-watt batteries. Solid-state thermoelectric power generators on flexible substrates[2, 4, 5] are attractive alternatives that eliminate recharging, replacement, and electrolyte leakage[6] issues in batteries. Efficient thermoelectric generators require materials with high figure-of-merit[7] $ZT = \frac{\alpha^2 \sigma T}{\kappa}$, which in turn requires a high Seebeck coefficient $\alpha$, high electric conductivity $\sigma$, low thermal conductivity $\kappa$ at operating temperature T, along with well-engineered p-n junctions and metal/thermoelectric interfaces[8-12].

Integrating high ZT inorganic thermoelectrics[13] with organic substrates is attractive for mechanically flexible high-efficiency thermoelectric power generators[14]. For instance, the high ZT of pnictogen chalcogenides[15-17] can be increased further by nanostructuring-induced decrease in $\kappa$, and maximizing power factor $\alpha^2\sigma$ via doping and alloying[17, 18] to manipulate the electronic structure. However, integrating these inherently inorganic brittle materials with organic substrates is a challenging task[19-22]. One strategy to circumvent cracking is to use thin film assemblies of inorganic nanostructures, which offer enhanced mechanical compliance and improved structural stability[23].

Thin film assemblies of n-type $Ag_2Te$[14, 24] and p-type $Sb_2Te_3$[15, 16] nanostructures exhibit high $\alpha^2\sigma$ near room temperature. Anisotropic 1-D and 2-D nanostructures often enhance ZT[20, 25] through decreased thermal conductivity $\kappa$ and high power factor from nanoscopic confinement effects. These findings have spurred studies of nanostructured thin films[15, 26] of $Ag_2Te$ and $Sb_2Te_3$ on flexible substrates (e.g., polyvinylidene fluoride –PVDF[20], nylon[21, 27], polyimide[28,29, 30], paper[31], and polytetrafluoroethylene[21]) by physical[32,33]/chemical[34] methods including vacuum filtration[20, 27], drop casting[35, 36], and inkjet/screen printing[2, 37]. For instance, $Ag_2Te$/nylon devices[27] yield high power factors $\alpha^2\sigma \sim 513$ μWm$^{-1}$K$^2$ and ~1.6 μW maximum power. Composites of $Sb_2Te_3$/MXene[38] and $Sb_2Te_3$/carbon nanotubes[39] and $Bi_2Te_3$/$Sb_2Te_3$ films integrated with paper[35, 36] yield ~40-45 mV at $\Delta T = 30$-50 °C, but the power output is low in the pW range.

Here, we demonstrate the performance of a thermoelectric power generator fabricated from microwave-synthesized $Ag_2Te$ nanowires and sulfur-doped $Sb_2Te_3$ nanoplates on PVDF. The morphology engineering using microwave synthesis is ~50- to 100-fold faster (e.g., <~ 10 minutes)[15, 16, 18, 40] than solvothermal[41-43], polyol[20, 44], and electrodeposition methods[45] (e.g., ~>24 hours)[41, 43]. By varying the sulfur content in p-$Sb_2Te_3$ (1-1.5 at.%), we achieved significant improvements in carrier concentration and power factor $\alpha^2\sigma$. A six-leg thermoelectric device comprised of alternating strips of n-$Ag_2Te$ and p-$Sb_2Te_3$, generated ~ 33.6 mV at $\Delta T \sim 50$ °C and a power of ~120 nW at $\Delta T \sim 30$ °C. These results highlight a scalable approach to fabricate thermoelectric power generators for harvesting low-grade waste heat.

## 2. Experimental Section

### 2.1. *Sb₂Te₃ nanoplate synthesis*

$Sb_2Te_3$ nanoplates were synthesized by a microwave solvothermal process described elsewhere[16]. $SbCl_3$, 95% pure thioglycolic acid (TGA), 95% pure 1,5-pentanediol, 99.8% pure 200-mesh Te powder, and tri-n-octylphosphine (TOP) obtained from Sigma Aldrich were used as-received. A homogeneous 0.15 mmol Te solution was obtained by dissolving Te powder in 2 ml TOP with 10 minutes sonication, followed by three cycles of ~ 900 W microwave exposure for ~1-2 minutes. The solution was then cooled for ~1-2 minutes. In parallel, 225 ml TGA was added to a 0.30 mmol SbCl$_3$ solution in 6.25 ml 1,5-pentanediol and sonicated for ~20 minutes. This TGA-SbCl$_3$ solution was mixed with the Te solution and continuously stirred for ~5 minutes prior to a 2-minute microwave exposure. Cooling the reaction mixture to room temperature yielded dark grey precipitates which were centrifuged, repeatedly washed with acetone and isopropanol, and dried overnight at 60 ℃.

### 2.2. Ag$_2$Te nanowire synthesis

Ag$_2$Te nanowires were synthesized by microwave-stimulation from as-received 97~98% pure NaOH and K$_2$TeO$_3$·H$_2$O, ~99.0% pure AgNO$_3$ and ethylene glycol, all from Sigma-Aldrich. First, a homogenous 4.95 mmol NaOH solution was prepared in 25 ml ethylene glycol by ultrasonication for ~ 60 minutes. In parallel, 0.5 mmol K$_2$TeO$_3$·H$_2$O was dissolved in ~15 ml ethylene glycol and ultrasonicated for ~ 30 minutes. Following this, ~1 mmol AgNO$_3$ was dissolved in ~ 5 ml ethylene glycol by continuous stirring for ~ 5 minutes, and mixed with the Te solution in an Erlenmeyer flask. The NaOH solution was added gradually to the Ag-Te solution mixture to obtain a black solution consisting of ~ 45 ml of 110 mM NaOH, 11 mM K$_2$TeO$_3$·H$_2$O, and 22 mM of AgNO$_3$.

This solution was exposed to repeated cycles of ~ 15-second microwave exposures followed by stirring. The first 10 microwave exposure cycles were separated by 10 seconds stirring intervals, the second 10 cycles were interspersed with 25 seconds stirring, and the last five cycles were spaced with 60 seconds stirring. The reaction temperature was maintained at 180–195 ℃, with the sample exposed to an estimated ~ 270 kJ of microwave energy. The black precipitates obtained were separated by centrifugation, washed thoroughly with acetone and isopropanol, and dried overnight at 60 ℃.

### 2.3. Device fabrication

Thin film assemblies of Sb$_2$Te$_3$ nanoplates and Ag$_2$Te nanowires were deposited on PVDF by vacuum filtration from 0.08 g powders of the respective materials dispersed in 20-ml ethanol solutions. These solutions were sonicated for ~10 minutes to ensure uniform nanoparticle dispersions. Vacuum filtration of the solution through a 47-mm PVDF membrane with 0.22 μm pores resulted in uniform films (Fig. 1) that were dried at 80 ℃ for ~10 minutes to remove solvent traces. The film was cold-pressed by gradually increasing the load from 0 to 7.5 metric tons. Alternate strips of 5 mm x 28 mm Sb$_2$Te$_3$/PVDF and Ag$_2$Te/PVDF structures were connected using Cu wires contacts capped with quick-dry Ag epoxy (SPI 04998-AB) applied using a needle to fabricate a 6-leg thermoelectric device.

### 2.4. Characterization and device performance testing

Output voltage and power characteristics were measured as a function of temperature gradient, ΔT, (Fig. S1a) applied using a hot plate and a water cooler aluminium block affixed with 100-μm-diameter E-type thermocouples. These measurements were carried out in a custom-built setup automated through a LabView interface to a Keithley source meter 2400 and a Keithley multimeter 2000 (Fig. S1b).

X-ray diffractograms (XRD) were obtained using a PANalytical instrument with a Cu Kα probe beam for phase identification. A field-emission Zeiss Supra 55 SEM operated at 5-20 keV was used to characterize nanostructure morphology, flexible films thickness and determine chemical composition from energy-dispersive X-ray (EDX) spectra obtained at 20 keV. X-ray photoelectron spectroscopy (XPS) measurements in a PHI 5000 Versaprobe system were used to characterize the surface chemistry. XPS analyses used MultiPak software (ULVAC-PHI, Inc.) to subtract background by the Shirley method, and corrected charging-related energy shifts with reference to the adventitious carbon peak.[46] A second order Savitzky-Golay filter with a 10-point window was used to smoothen the oxidized S 2p spectra.

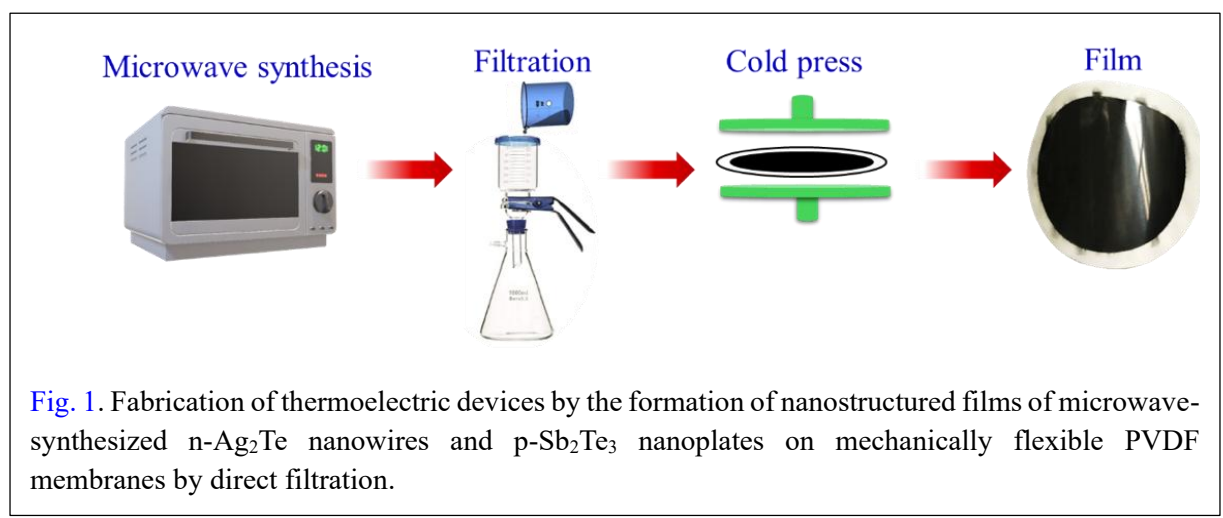

Fig. 1. Fabrication of thermoelectric devices by the formation of nanostructured films of microwave-synthesized n-Ag$_2$Te nanowires and p-Sb$_2$Te$_3$ nanoplates on mechanically flexible PVDF membranes by direct filtration.

Electrical conductivity σ was determined with a Ossila T2001A four-point-probe station. The Seebeck coefficient α was determined from the Seebeck voltage measured with a Keithley 2001 multimeter[47] in a custom-built setup with the temperature gradient created by Peltier modules and measured by K-type thermocouples. Charge carrier concentration, n and mobility, µ were calculated from room-temperature Hall measurements in an apparatus with a ~ 0.485 T permanent magnet. Hall voltages were measured in van der Pauw geometry using a Keithley 2100 multimeter for a 10 mA current delivered by a Keithley 2400 source. The effect of mechanical deformation of the nanostructured films on PVDF on power output was tested by compressively flexing the film in an Instron 5843 system.

## 3. Results and discussion

### 3.1. Nanocrystal structure, morphology and composition

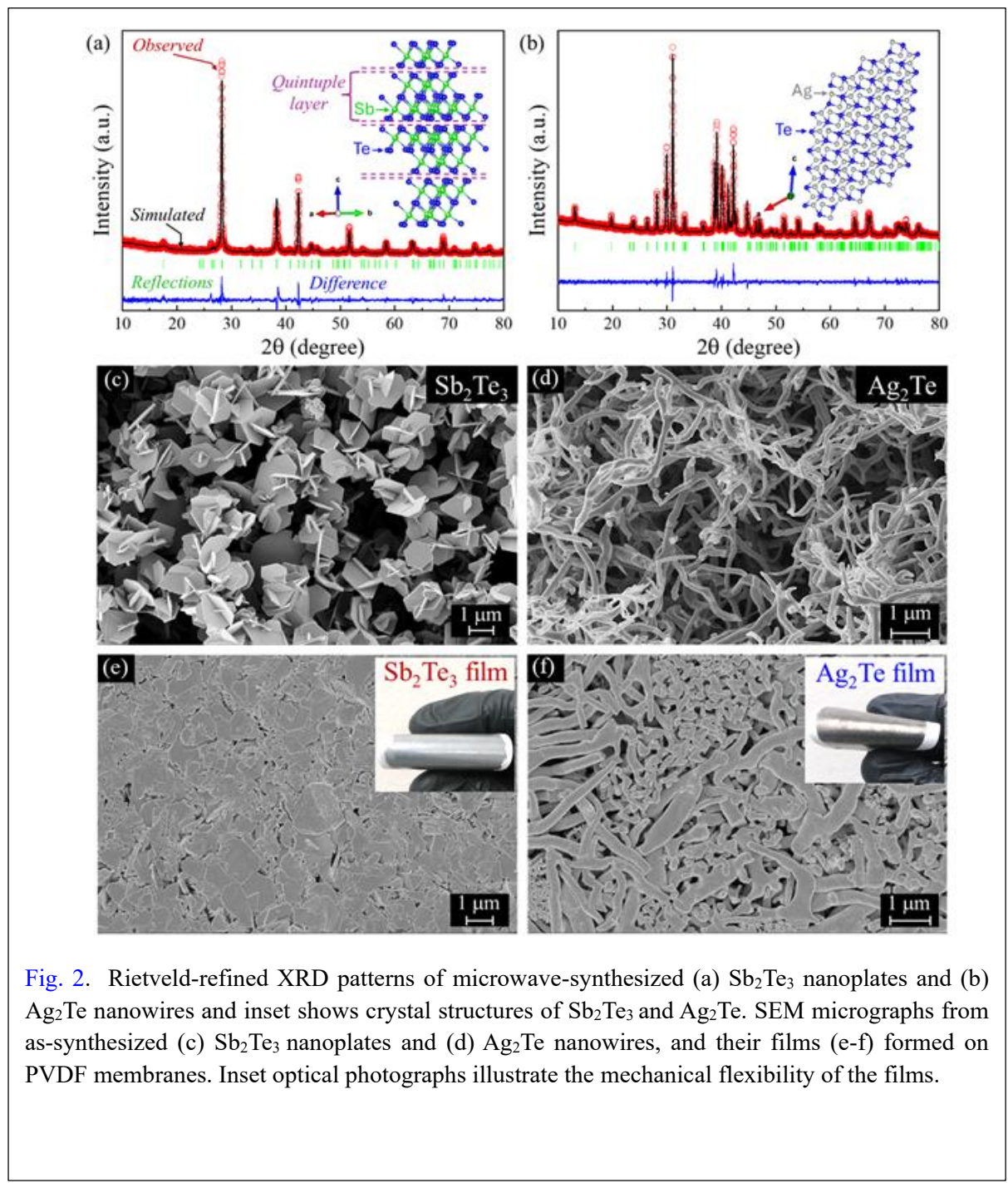

 Rietveld-refined XRD patterns of microwave-synthesized (a) $Sb_2Te_3$ nanoplates and (b) $Ag_2Te$ nanowires and inset shows crystal structures of $Sb_2Te_3$ and $Ag_2Te$. SEM micrographs from as-synthesized (c) $Sb_2Te_3$ nanoplates and (d) $Ag_2Te$ nanowires, and their films (e-f) formed on PVDF membranes. Inset optical photographs illustrate the mechanical flexibility of the films.

X-ray diffractograms show the presence of rhombohedral $Sb_2Te_3$ (JCPDS No. 15-874 with R$\bar{3}$m space group), and monoclinic $Ag_2Te$ (JCPDS No. 34-142, with P$2_1$/c space group). The positions of the sharp peaks (Fig. 2a-b) from both materials agree with the JCPDS cards, indicating that microwave synthesis yields high quality and high purity materials. Unit cell dimensions obtained from Rietveld refinement analyses are listed in supplementary information Table S1. SEM images show hexagonal $Sb_2Te_3$ nanoplates and branched $Ag_2Te$ nanowires (Fig. 2c-d) with monomodal size distributions (see supplementary Fig. S2). EDX reveals uniform elemental distributions in $Sb_2Te_3$ and $Ag_2Te$ nanostructures (see Fig. S3), highlighting compositional homogeneity. $Ag_2Te$ nanostructure shape is sensitive to microwave cycling (Fig. S4). Pulsed microwave heating with on/off cycles yields $Ag_2Te$ nanowires (Fig. S4a-b), in contrast to near-spherical Ag nanoparticles obtained by continuous heating (Fig. S4c-d). Microwave cycling likely provides sufficient time for $TeO_3^{2-}$ reduction which is known to favor nanowire formation.[48]

$Ag_2Te$ nanostructures can be envisioned to form in five steps as follows. Ethylene glycol dehydration (1) yields acetaldehyde which reduces $Ag^+$ to produce Ag metal (2). Potassium tellurite ionization (3) yields $TeO_3^{2-}$ ions which are reduced to Te metal (4), which combines with Ag to form $Ag_2Te$ (5). The chemical reactions are as follows,

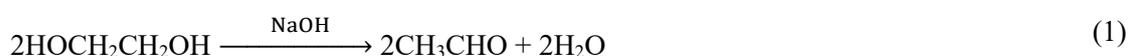

$$2HOCH_2CH_2OH \xrightarrow{\text{NaOH}} 2CH_3CHO + 2H_2O \tag{1}$$

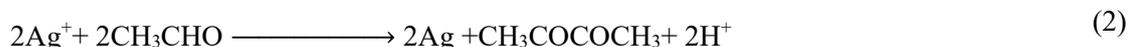

$$2Ag^+ + 2CH_3CHO \longrightarrow 2Ag + CH_3COCOCH_3 + 2H^+ \tag{2}$$

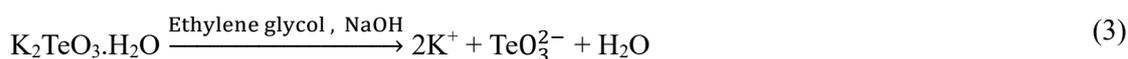

$$K_2TeO_3.H_2O \xrightarrow{\text{Ethylene glycol , NaOH}} 2K^+ + TeO_3^{2-} + H_2O \tag{3}$$

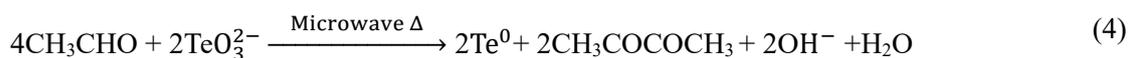

$$4CH_3CHO + 2TeO_3^{2-} \xrightarrow{\text{Microwave } \Delta} 2Te^0 + 2CH_3COCOCH_3 + 2OH^- + H_2O \tag{4}$$

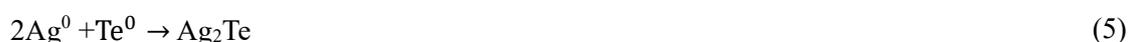

$$2Ag^0 + Te^0 \rightarrow Ag_2Te \tag{5}$$

### *3.2. Effects of surfactants on phase formation and morphology*

Microwave synthesis of $Sb_2Te_3$ is sensitive to precursors and surfactant concentrations (Figs. 3a-f), which provides significant scope to control $Sb_2Te_3$ nanostructure morphology and Te heterostructuring. The use of trioctylphosphine (TOP) surfactant yields $Sb_2Te_3$ (Figs. 3a-b). While 2 ml TOP yields well-formed hexagons, increasing TOP to 5 ml leads to the decoration of Te fins at platelet edges[40] (Figs. 3a-b). Replacing TOP with trioctylphosphine oxide (TOPO) produces $Sb_2Te_3$ with deformed hexagons morphology rough surface along with elemental Te (see Figs. S5a-b). Increasing the pentanediol content has a similar effect as increasing TOP; bare hexagon-shaped $Sb_2Te_3$ platelets give way $Sb_2Te_3$-Te heterostructures with Te fins (Figs. 3c-d) reminiscent of prior works[15, 40]. Low thioglycolic acid (TGA) concentrations (e.g., ~0.14 M) yield poorly-defined nanoplates due to non-uniform nucleation and growth (Fig. 3e). Hexagon-shaped $Sb_2Te_3$ nanoplates (Fig. 3f) are formed with 0.5 M TGA, in agreement with earlier studies[40]. Besides shape control via selective passivation of $Sb_2Te_3$ crystal facets by thiol moieties[49], TGA also facilitates sulfur doping[15]. Absolute precursor amounts also influence the $Sb_2Te_3$ morphology (Figs. S6a-b).

### 3.3. Partial oxidation

Core-level spectra from nanostructured $Sb_2Te_3$ and $Ag_2Te$ films reveal surface oxidation (Figs. 4a-b; Fig. S8). The high binding energy Sb $3d_{3/2}$ and $3d_{5/2}$ peak components at ~538.2 eV and ~528.8 eV from $Sb_2Te_3$ films correspond to $Sb_2Te_xO_{3-x}$ (Fig. 4a). This is corroborated by the high-binding energy Te $3d_{5/2}$ – Te $d_{3/2}$ components at ~576.4 eV and 586.7 eV, respectively indicating $TeO_x$ (Fig. 4b). The presence of both unoxidized and oxidized Te components indicate partial surface oxidation. The 170 eV $SO_x$ peak (Fig. 4b inset) confirms that the $Sb_2Te_3$ nanoplates are sulfur-doped. Thus TGA is a triple-agent for sulfur-doping and nanostructure-shaping[16]. Films of $Ag_2Te$ nanowires show overlapping bands of Te $3d_{5/2}$ at ~ 572.8 eV and Ag $3p_{3/2}$ at ~ 573.2 eV (Fig. 4c-d), indicating partial surface oxidation with a nominal $AgTe_{0.1}O_{0.2}C_{0.4}$ stoichiometry. Fig. S9 shows TGA-induced increase in sulfur bands from $Sb_2Te_3$ films on PVDF film.

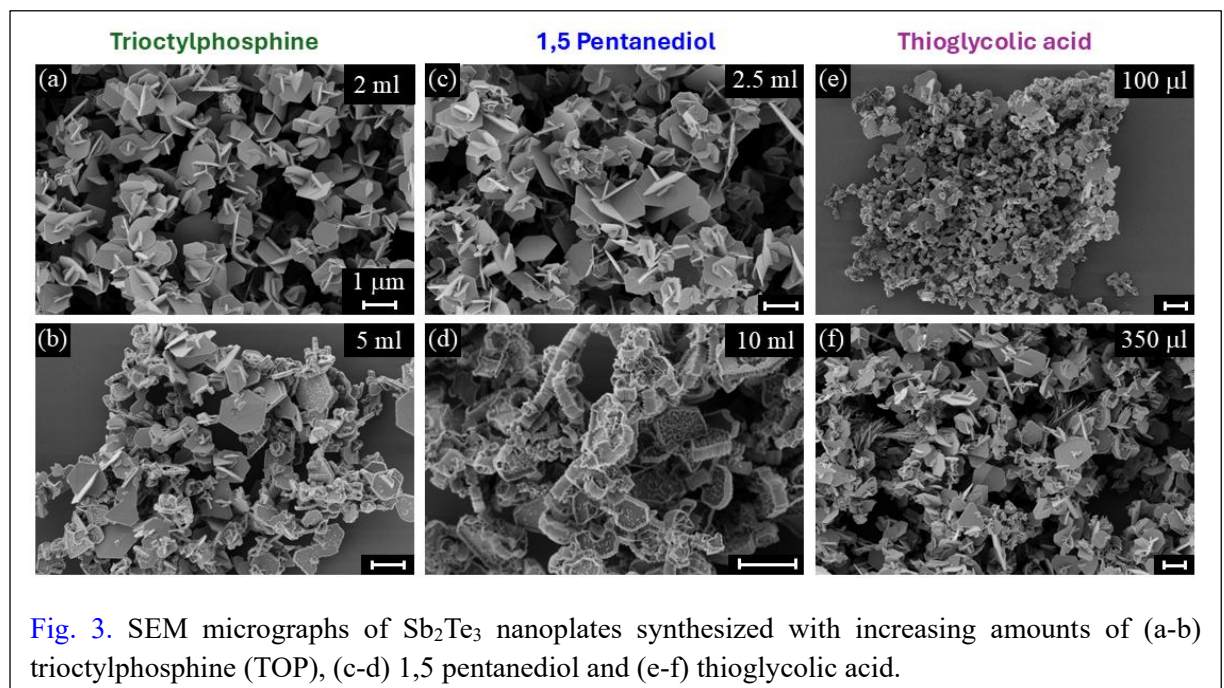

| **Trioctylphosphine** | **1,5 Pentanediol** | **Thioglycolic acid** |
|---|---|---|
| (a) 2 ml | (c) 2.5 ml | (e) 100 µl |
| (b) 5 ml | (d) 10 ml | (f) 350 µl |

Fig. 3. SEM micrographs of $Sb_2Te_3$ nanoplates synthesized with increasing amounts of (a-b) trioctylphosphine (TOP), (c-d) 1,5 pentanediol and (e-f) thioglycolic acid.

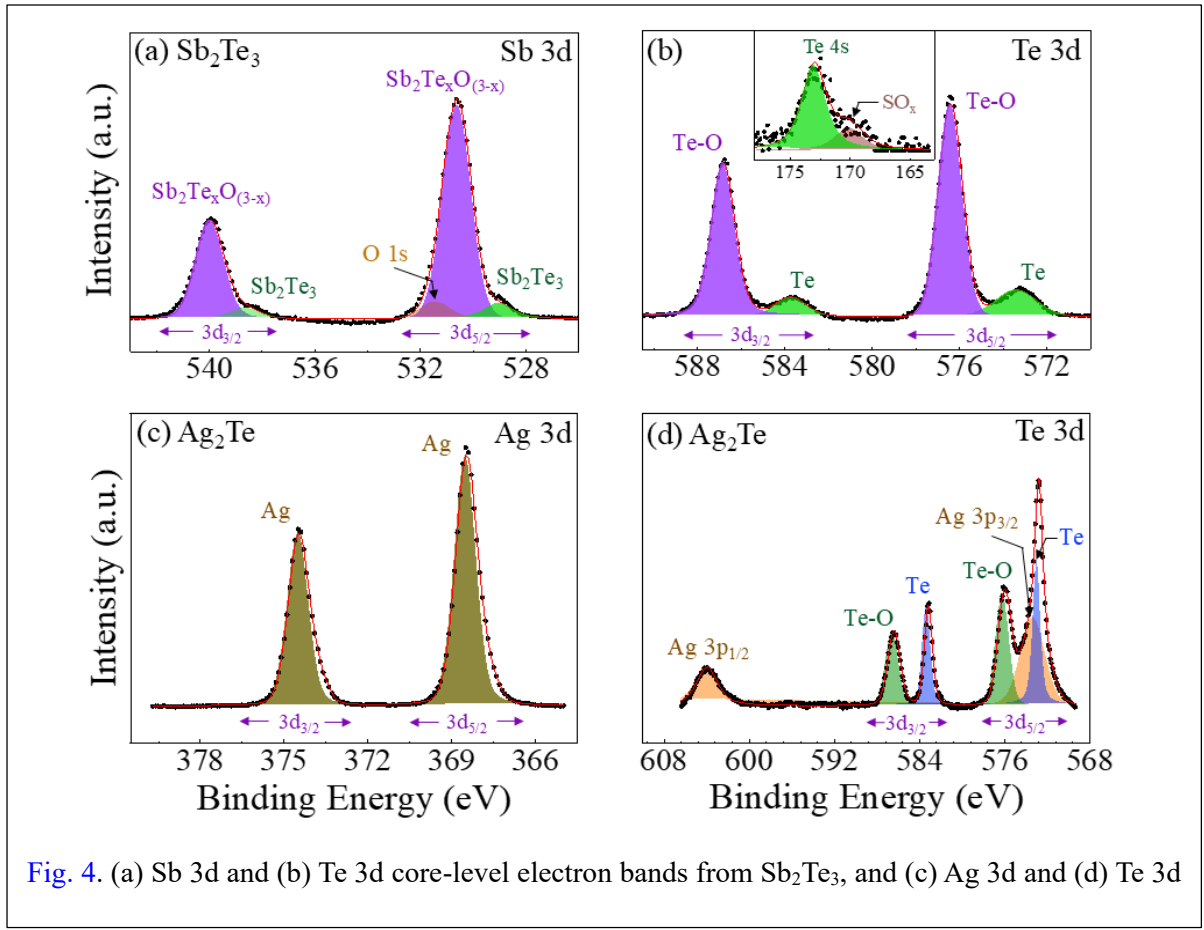

**Fig. 4.** (a) Sb 3d and (b) Te 3d core-level electron bands from Sb$_2$Te$_3$, and (c) Ag 3d and (d) Te 3d

### 3.4. Thermoelectric properties of films of Ag$_2$Te and Sb$_2$Te$_3$ nanostructure assemblies on PVDF

Figs. 5a-b summarizes the electrical conductivity $\sigma$ measured and Seebeck coefficient $\alpha$ determined from Ag$_2$Te and Sb$_2$Te$_3$ films. Ag$_2$Te films show the highest $\sigma = 50000 \pm 990$ Sm$^{-1}$ and a power factor $\alpha^2\sigma = 150 \pm 55.1$ $\mu$Wm$^{-1}$K$^{-1}$ and $\alpha = -55 \pm 10$ $\mu$VK$^{-1}$ indicating n-type conduction. Sb$_2$Te$_3$ films with both sulfur doping levels x$_S$=1 at.% and 1.5 at.%. show $\alpha$>0, indicating p-type behavior. Higher sulfur doping results in a fivefold higher $\sigma = 1800 \pm 3.4$ Sm$^{-1}$, a ~25% higher $\alpha$ ~130 $\pm$18 $\mu$VK$^{-1}$, and an eightfold higher power factor $\alpha^2\sigma = 30 \pm 8.3$ $\mu$Wm$^{-1}$K$^{-1}$. These enhancements correlate with twenty-fold higher hole concentration h = 8.80 x 10$^{19}$ cm$^{-3}$ for x$_S$ =1.5 at.% than h = 4.30 x 10$^{18}$ cm$^{-3}$ for x$_S$ = 1 at.%

indicated by room-temperature Hall measurements (Table S2). Thus, the sulfur doping level is a key factor to tailor electronic transport and thermoelectricity in Sb₂Te₃, consistent with prior works[16].

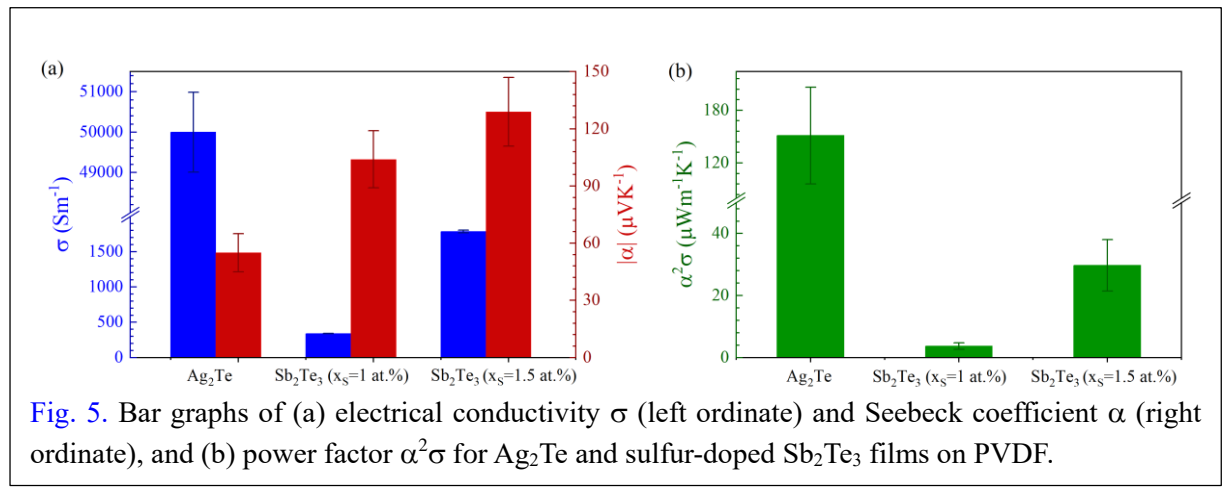

Fig. 5. Bar graphs of (a) electrical conductivity σ (left ordinate) and Seebeck coefficient α (right ordinate), and (b) power factor α²σ for Ag₂Te and sulfur-doped Sb₂Te₃ films on PVDF.

### 3.5. Thermoelectric device fabrication and power generation characteristics

A six-leg p-n junction thermoelectric device is fabricated by integrating alternating PVDF strips with Ag₂Te and 1.5 at.% sulfur-doped Sb₂Te₃ films onto a polyimide substrate (Fig. 6a). Mechanical bending significantly increases the device electrical resistance (see Fig. 6b) linearly with apex bending angle for $0° < \theta < \sim 50°$. At higher angles (e.g., $\theta \sim> 60°$) the resistance increases at a lower rate, yielding a cumulative resistance increase of ~12.5% at the highest angle $\theta \sim 99.7°$. No discernible macroscale changes in the film are revealed by SEM images, suggesting that the bending-induced resistance changes likely arise from small perturbations in between nanostructures relative to each other, the PVDF membrane and the metal contacts.

Repeated bending with $\theta_{min} = 0°$ and $\theta_{max} \sim 80°$ shows a monotonic increase in resistance with increasing number of bending cycles $n_{bending}$ (Fig. S10). The ~124% increase in resistance for $n_{bending} = 200$ cycles is larger than that reported for Zn-doped Sb₂Te₃ films deposited Sb/Zn magnetron sputtering and Te evaporation followed by annealing[50]. This large change is attributable to inter-nanostructure microcracking in the films (see inset in Fig. S10) that disrupt low-conductance network paths[22, 51]. Including a low-resistance binder[52] could potentially counter this behavior.

Our six-leg devices show an output voltage $V_{OC}$ = 33.6 mV for $\Delta T$ = 50 °C, and a maximum power output $P_{OC}$ =120 nW at $\Delta T$ = 30 °C (Figs. 6c-d). The estimated voltage and power outputs per gram, namely, are $V_m \sim$ 396 mV/g for $\Delta T \sim$ 50 °C and $P_m \sim$ 1.24 μW/g for $\Delta T$ ~30 °C (Table S3) surpass reported values for $Sb_2Te_3$/membrane and paper-based devices[35, 36], but are inferior to the state-of-the-art thermoelectric devices (see supplementary Table S4). A human finger touch produced ~ 2.6 mV from our thermoelectric device (See supplementary movie S1). These findings demonstrate that films from microwave-synthesized nanostructures on PVDF are promising as heat-harvesting power generators. The relatively high internal resistance of ~ 850 Ω in our devices are likely due to surface oxidation and organic capping of the nanostructures, which could be addressed through post-deposition surface oxide reduction by hydrazine, acid treatments to remove organics[40, 52], and thermal annealing.

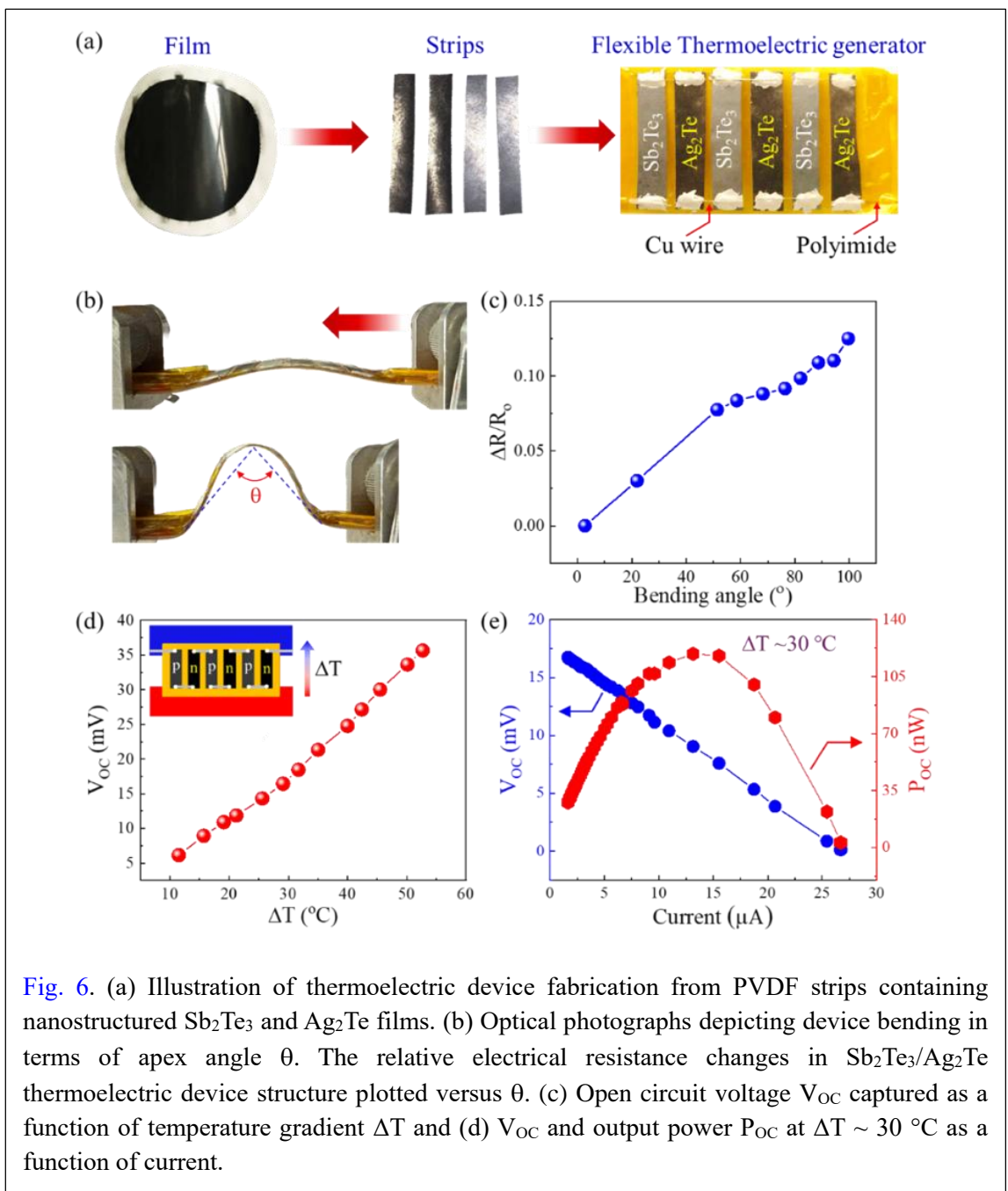

Fig. 6. (a) Illustration of thermoelectric device fabrication from PVDF strips containing nanostructured $Sb_2Te_3$ and $Ag_2Te$ films. (b) Optical photographs depicting device bending in terms of apex angle θ. The relative electrical resistance changes in $Sb_2Te_3$/$Ag_2Te$ thermoelectric device structure plotted versus θ. (c) Open circuit voltage $V_{OC}$ captured as a function of temperature gradient $\Delta T$ and (d) $V_{OC}$ and output power $P_{OC}$ at $\Delta T \sim$ 30 °C as a function of current.

## 4. Conclusions

Ag-rich $Ag_2Te$ nanowires and sulfur-doped $Sb_2Te_3$ nanoplates were produced by rapid microwave synthesis, yielding gram-scale quantities of high-quality materials within minutes. Sulfur doping in the 1~1.5 at.% range in $Sb_2Te_3$ leads to an eightfold power factor increase driven by a fivefold improvement in electrical conductivity and a ~1.25-fold increase in the Seebeck coefficient. Six-leg p-n junction devices fabricated from nanostructured films of $Ag_2Te$ and sulfur-doped $Sb_2Te_3$ on PVDF exhibit promising characteristics such as a maximum voltage of ~33.6 mV for $\Delta T$= 50 °C and a maximum power of ~120 nW at $\Delta T$=30°C. These results highlight the promise of rapid synthesis of high-quality nanomaterials and their utility for thermoelectric power generation for flexible electronics devices. Process optimization and engineering inter-nanostructure interactions during bending should pave way for efficient, lightweight, and sustainable energy harvesting technologies.

**Conflicts of interest:** There are no conflicts to declare.

**Data availability**

The data supporting this article have been included as part of the supplementary information (SI). Supplementary information is available.

**Acknowledgements**

A.S. and A.K. acknowledge research facility at IIT Mandi and funding from Department of Science and Technology, New Delhi, India. P. E. acknowledges funding from the Knut and Alice Wallenberg Foundation through the Wallenberg Academy Fellows program (grant no. KAW 2020.0196, and the Swedish Research Council under project grants no 2021-03826, and 2025-03680. G.R. acknowledges funding support from the US National Science Foundation grant CMMI 2135725 through the BRITE program, the Empire State Development's Division of Science, Technology and Innovation Focus Center at RPI, C210117, Wallenberg Initiative in Materials Science for Sustainability (WISE) funded by the Knut and Alice Wallenberg Foundation, and a Swedish Research Council grant under VR project 2024-04996.

# Morphology-engineered nanostructured silver- and antimony-telluride films for flexible thermoelectric generators

Ankit Kashyap[a,b], Conner Wallace[b], Geetu Sharma[b], Collin Rowe[b], Mahima Sasikumar[b], Niraj Kumar Singh[d], Per Eklund[d,e], Theodorian Borca-Tasciuc[c], Ganpati Ramanath[b,d,e*], and Ajay Soni[a*]

[a]School of Physical Sciences, Indian Institute of Technology Mandi, 175005, Himachal Pradesh, India.

[b]Rensselaer Polytechnic Institute, Materials Science & Engineering Department, Troy, NY 12180, USA.

[c]Rensselaer Polytechnic Institute, Department of Mechanical Engineering, Troy, NY 12180, USA.

[d]Inorganic Chemistry, Ångström Laboratory, Uppsala University, SE-751 21 Uppsala, Sweden.

[e]Wallenberg Initiative Materials Science for Sustainability (WISE), Uppsala University, Sweden.

*Corresponding authors: ganapr@rpi.edu, ganpati.ramanath@kemi.uu.se and ajay@iitmandi.ac.in

The data and explanations provided below supplement the narrative and figures in the main manuscript.

## 1. Custom-built setup used for flexible thermoelectric device measurements

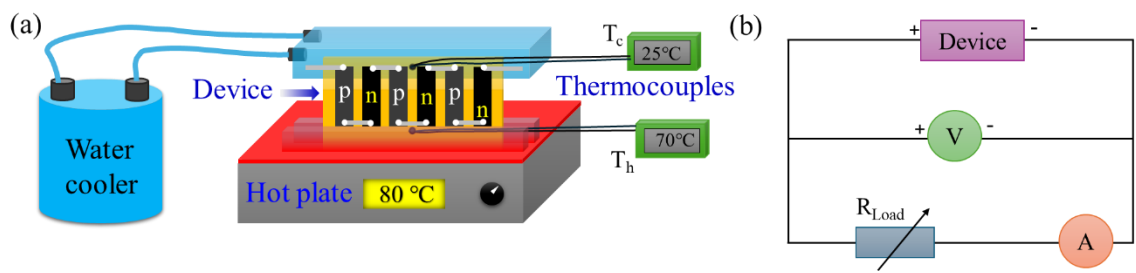

**Figure S1.** Schematic of the (a) custom setup used for the measurement and (b) circuit diagram of measurement.

Performance evaluation of flexible thermoelectric generators is done using a custom-built performance measurement system. Lateral temperature gradient is applied using a hot plate and a water-cooled aluminium block on the hot and cold side respectively. The temperature on cold and hot side was measured by attaching a E-type thermocouple of diameter (100 μm) on the device. Output voltage as a function of ΔT is measured using Keithley multimeter 2000 and for the output power measurement variable resistance is connected in series to the source meter as shown in the circuit diagram Figure S1. The power output measurement of flexible thermoelectric generator was measured using Keithley source meter 2400 (measure the current output) and Keithley multimeter 2000 (measure the output voltage) through a LabVIEW program.

## 2. Lattice parameters from x-ray diffraction

| Material | a (Å) | b (Å) | c (Å) | Lattice volume (Å³) |
|---|---|---|---|---|

| Sb$_2$Te$_3$ | 4.2720 | 4.2720 | 30.3953 | 480.4013 |
| Ag$_2$Te | 8.1671 | 4.4697 | 8.9795 | 271.1747 |

Table S1. Lattice parameters obtained using Rietveld refinement of Sb$_2$Te$_3$ and Ag$_2$Te.

Lattice parameters of Sb$_2$Te$_3$ and Ag$_2$Te were extracted through Rietveld refinement of the powder X-ray diffraction data. Refinements were performed using *FullProf* software employing a pseudo-Voigt profile function and polynomial background fitting. The refined lattice parameters, lattice volumes, are summarized in Table S1.

### 3.  Size distribution of Sb$_2$Te$_3$ nanoplates and Ag$_2$Te nanowires

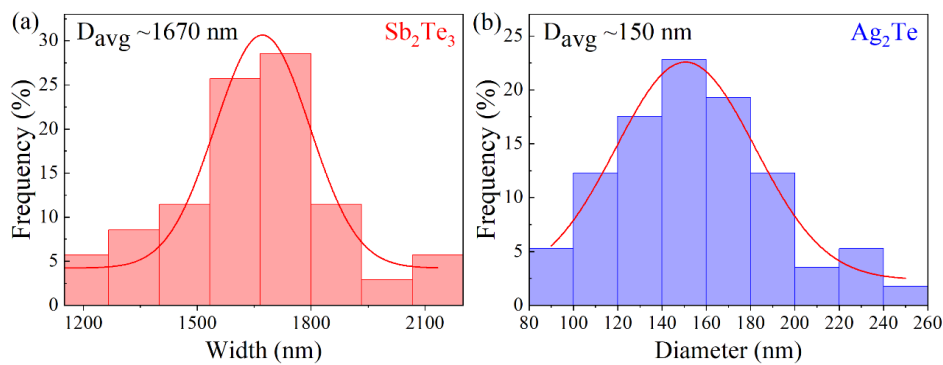

Figure S2.  Size distribution of microwave synthesized (a) Sb$_2$Te$_3$ nanoplates and (b) Ag$_2$Te nanowires.

Size distribution of microwave synthesized Sb$_2$Te$_3$ nanoplates and Ag$_2$Te nanowires was measured from scanning electron microscopy (SEM) images using ImageJ software. The average size of Sb$_2$Te$_3$ nanoplatelets and diameter of Ag$_2$Te nanowires obtained from SEM images are ~1670 nm ± 208 nm and ~150 nm ± 37 nm, respectively.

## 4. Elemental distributions in nanostructures

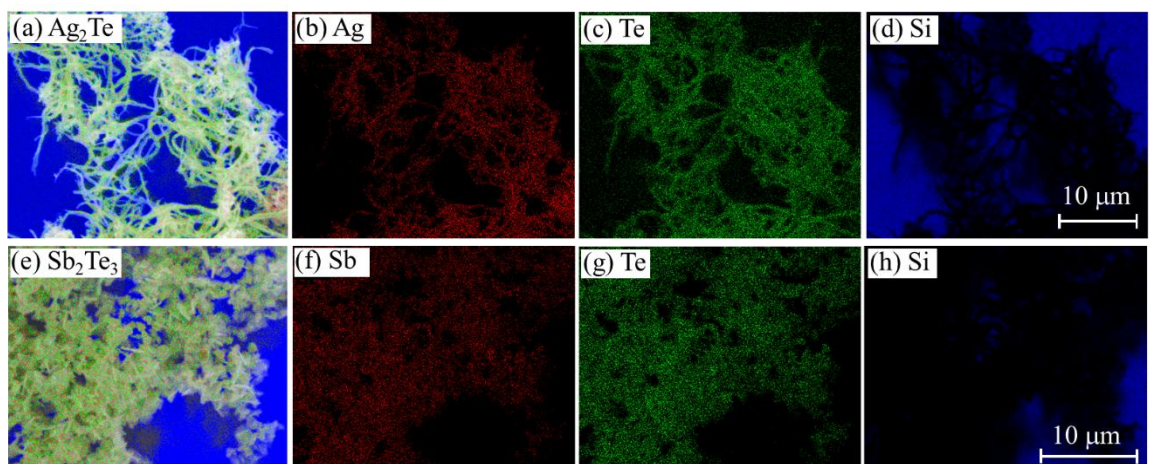

 Elemental mapping of Ag$_2$Te nanowires: (a) combined elemental map, (b) Ag distribution, (c) Te distribution, and (d) Si distribution. Elemental mapping of Sb$_2$Te$_3$ nanoplates: (e) combined map, (f) Sb distribution, (g) Te distribution, and (h) Si distribution.

A field-emission Zeiss Supra 55 SEM equipped with an energy dispersive X-ray (EDX) detector was used to collect elemental maps and to examine the morphology of the nanostructures. All EDX measurements were performed at an accelerating voltage of 20 keV.

## 5. Ag$_2$Te nanowires formation dependence of reaction conditions

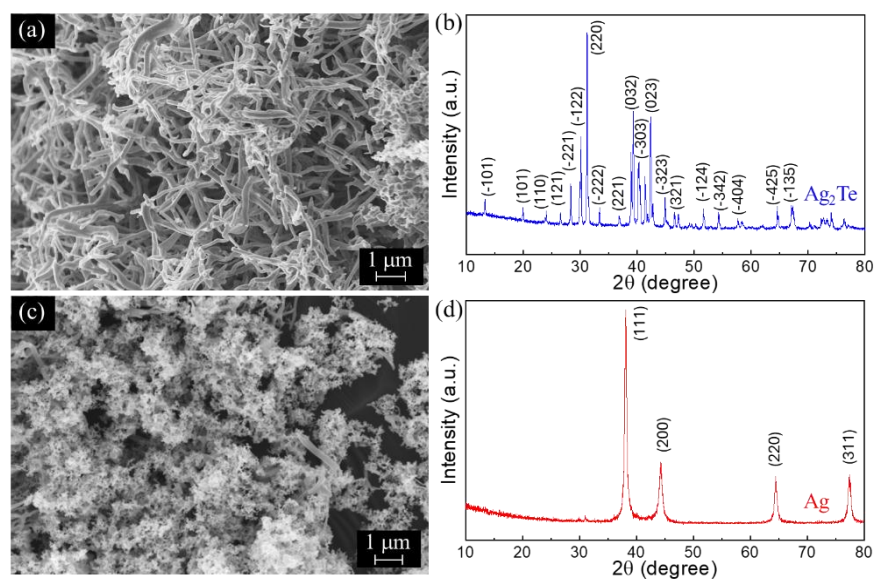

**Figure S4.** Scanning electron micrographs of (a) pulsed microwave heated reaction, (b) x-ray diffraction patterns of Ag$_2$Te (c) Ag nanoparticles and (d) x-ray diffraction patterns of Ag.

The formation of Ag$_2$Te nanowires was found to depend strongly on the microwave-heating conditions. Cyclic microwave heating, where the power is periodically switched on and off, promotes the growth of Ag$_2$Te nanowires. In contrast, continuous (non-cycled) microwave heating results in the formation of Ag nanoparticles instead, as shown in Figure S4.

## 6. X ray diffraction patterns of Sb₂Te₃ synthesized using TOP and TOPO

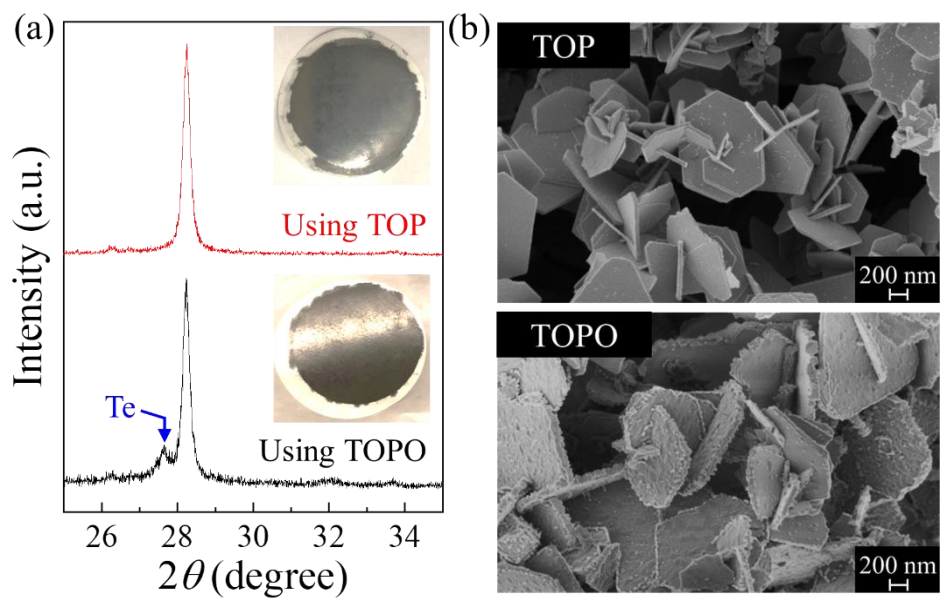

 (a) X-ray diffraction patterns of Sb₂Te₃ synthesized using TOP (red) and TOPO (blue), highlighting structural differences, with inset images showcasing the corresponding films. (b) Scanning electron microscopy images showing Sb₂Te₃ synthesized using TOP with smooth hexagon, and TOPO with rough hexagonal morphology.

## 7. Effect of precursor amounts on Sb₂Te₃ morphology

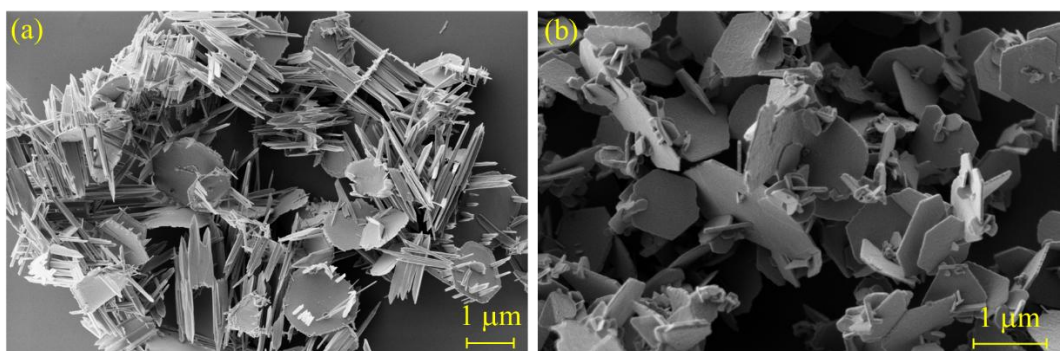

**Figure S6.** SEM images showing the effect of precursor amounts on Sb₂Te₃ morphology.

(a) Heterostructure hexagons with rod-like edges formed at low precursor amounts (0.04 mmol SbCl₃ in 6.25 ml 1,5-pentanediol, 0.08 mmol Te in 2 ml TOP). (b) No rods observed at higher amounts (0.15 mmol SbCl₃ and 0.30 mmol Te).

## 8. Cross-section SEM images of Ag₂Te and Sb₂Te₃ films

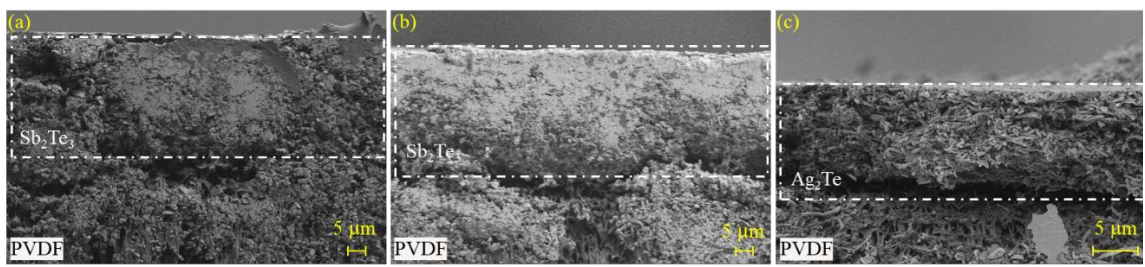

Figure S7. Cross-section SEM images of flexible films of (a) $Sb_2Te_3$ ($x_S$ = 1.5 at.%)1 (b) $Sb_2Te_3$ ($x_S$ = 1 at.%) and (c) $Ag_2Te$.

Cross-sectional measurements were performed on freshly fractured films using a Zeiss Supra 55 field-emission SEM. Obtained micrographs were used to estimate the thickness of films.

## 9. X-ray photoelectron spectroscopy $Ag_2Te$ and $Sb_2Te_3$ films

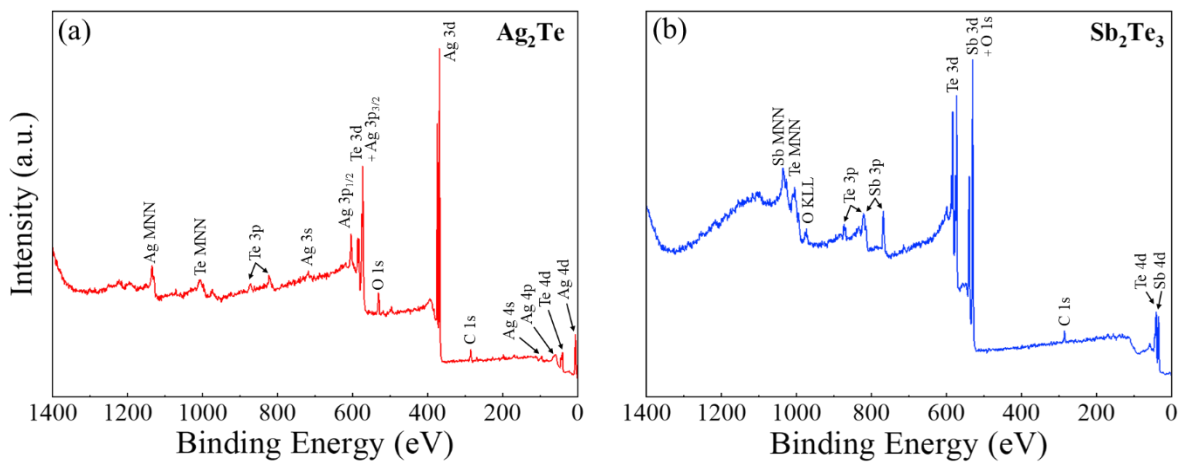

Figure S8. X-ray photoelectron spectroscopy survey scan of (a) $Ag_2Te$ and (b) $Sb_2Te_3$ film.

X-ray photoelectron spectroscopy (XPS) measurements in a PHI 5000 Versaprobe system were used to characterize the surface chemistry of $Ag_2Te$ and $Sb_2Te_3$ films. Al Kα X-rays were used as a source on 100 μm spot size. XPS analyses were performed using MultiPak software (ULVAC-PHI, Inc.) to subtract background by the Shirley method, and corrected charging-related energy shifts with reference to the adventitious carbon peak.

## 10. X-ray photoelectron spectroscopy of $Sb_2Te_3$ films

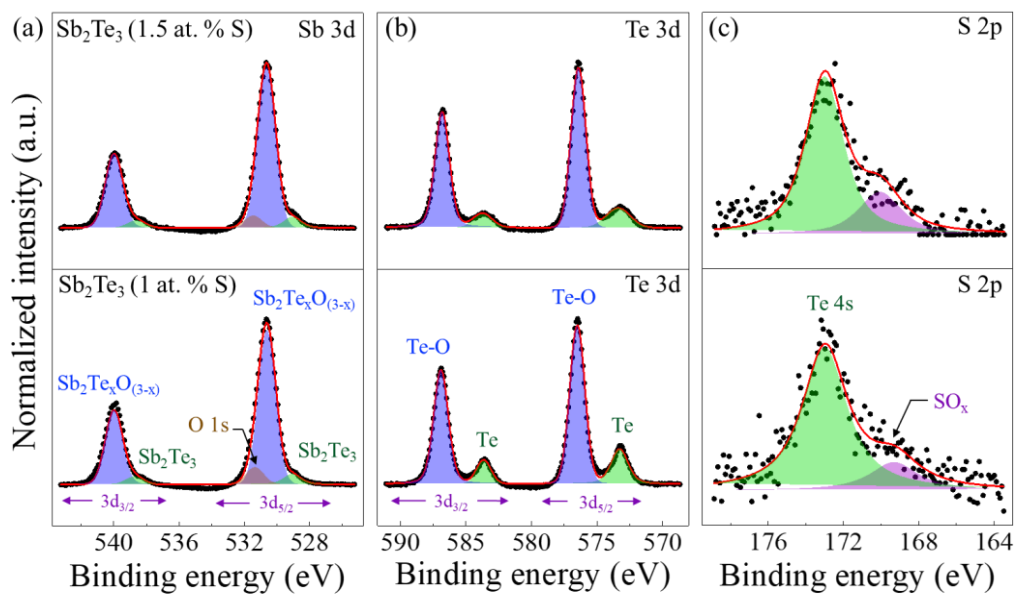

Figure S9. Comparison of x-ray photoelectron spectroscopy survey scan of (a) Sb 3d, (b) Te 3d and (c) S 2p for oxidised $Sb_2Te_3$ film with 1.5 at. % sulfur and 1 at. % sulfur.

## 11. Relative electrical resistance change with bending cycles

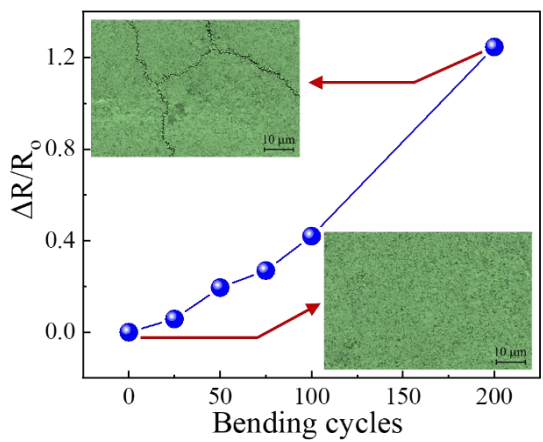

 The relative electrical resistance changes with the number of bending cycles for bending between $\theta_{min}=0°$ and $\theta_{max}\sim80°$ (inset showing SEM micrographs with highlighted crack area)

The electrical resistance as a function the bending cycle change was performed using a an Instron 5843 system by connecting ends of the flexible thermoelectric generator to a Keithley 2000 multimeter. SEM micrographs of the $Sb_2Te_3$ film on the flexible thermoelectric generator taken before and after bending the device films with the crack regions clearly highlighted and color-enhanced to improve visibility.

## 12. Hall measurement data of Ag₂Te and Sb₂Te₃ films

| Sample | n (cm$^{-3}$) | μ (cm$^2$V$^{-1}$s$^{-1}$) |
|---|---|---|
| $Sb_2Te_3$ ($x_S = 1$ at.%) | 4.30E+18 | 4.95E+00 |
| $Sb_2Te_3$ ($x_S = 1.5$ at.%) | 8.80E+19 | 1.27E+00 |
| $Ag_2Te$ | 4.21E+21 | 7.42E-01 |

Table S2. Hall measurement data of $Sb_2Te_3$ and $Ag_2Te$ films.

Hall measurements in an apparatus with a $\sim 0.485$ T permanent magnet. Hall voltages were measured in van der Pauw geometry using a Keithley 2100 multimeter for a 10 mA current delivered by a Keithley 2400 source.

## 13. Estimated output voltage and output power per gram

| Device | Voltage output per gram (50 K) | Power output per gram (30 K) |
|---|---|---|
| $Sb_2Te_3/Ag_2Te$ (6 leg device) | ~ 396 mV/g | 1.24 μW/g |

Table S3. Estimated voltage output and power output per gram at temperature difference of 50 K and 30 K respectively.

Voltage and power output per gram were calculated based on the measured device voltage and power output, normalized by the total mass of active material used in fabricating the device.

## 14. Comparison of flexible device performance with the literature

| Material | Number of legs | Power (nW) | ΔT (K) | Ref. |
|---|---|---|---|---|
| $Sb_2Te_3/Ag_2Te$ | 6 | ~120 | ~30 | This work |
| $Sb_2Te_3/Bi_2Te_3$ | 6 | ~ 0.35 | ~30 | [1] |
| $Bi_{0.5}Sb_{1.5}Te_3/$ $Bi_2Se_{0.3}Te_{2.7}$ | 10 | 10 | ~35 | [2] |
| $Bi_2Te_{2.8}Se_{0.2}$ | 5 | 1400 | ~30 | [3] |
| $Sb_2Te_3/MXene$ | 7 | 2000 | ~30 | [4] |

**Table S4**. Comparison of device performance with prior relevant reports.